\documentclass[twocolumn,prl,float]{revtex4}

\usepackage{graphicx} 
\usepackage{float}
\usepackage{mhchem} 
\usepackage{siunitx}
\usepackage{lipsum} 
\usepackage{epsfig}
\usepackage{ragged2e}
\usepackage{hyperref}
\usepackage{amsmath,amsfonts,amsthm,bm}
\usepackage{xcolor}

\begin{document}

%\title{Atomic-Level Insights Into \ce{TiO2-H2O} Interface Under Full Hydration} 

%\title{Hydrophobic signature on \ce{TiO2} nanoparticles in liquid water reveled by atomic-resolved interfacial protons.   }
%\title{Hydrophobic-hydrophilic nature of \ce{TiO2}  nanoparticles in liquid water}

\author{Lorenzo Agosta}

\affiliation{Department of Chemistry, $\AA$ngstr\"{o}m Laboratory, Uppsala University, 751 21 Uppsala, Sweden}

\email{lorenzo.agosta@kemi.uu.se}

%\author{Daniel Arismendi}

%\affiliation{Department of Chemistry, $\AA$ngstr\"{o}m Laboratory, Uppsala University, 751 21 Uppsala, Sweden}

\author{Mikhail Dzugutov}

\affiliation{Department of Chemistry, $\AA$ngstr\"{o}m Laboratory, Uppsala University, 751 21 Uppsala, Sweden}

\author{Wim Briels}

\affiliation{Faculty of Science and Technology, University of Twente, Netherlands}

\author{Kersti Hermansson}

\affiliation{Department of Chemistry, $\AA$ngstr\"{o}m Laboratory, Uppsala University, 751 21 Uppsala, Sweden}

%\author{authors}

%\newcommand{\EGB}[2]{}

%\title{Hydrophobic signature on \ce{TiO2} nanoparticles in liquid water}

%\title{What is the water structure on \ce{TiO2} nanoparticles in liquid water?  }
%\title{Water-induced hydrophobicity on \ce{CeO2} surface.}

%\title{Enhanced water diffusion on a hydrophilic \ce{CeO2} surface}
\title{The entropic origin of the  enhancement  of liquid diffusion at the confining wall.}

%hydrophilic, hydrophobic, diffusion

\begin{abstract}

We report a molecular dynamics simulation investigating the  dynamics of a simple liquid in the proximity to a non-interacting smooth confining wall. A strong enhancement of the liquid diffusion is observed within the layers adjacent to the wall. We present an analysis of these results  in terms of the scaling law earlier reported by one of us that relates the liquid diffusion rate to the excess entropy.   It is  demonstrated that this scaling law can successfully account for  the  observed diffusion  enhancement in the liquid  near to the confining wall. We show  that  the  proximity of a confining wall results in the  decrease of (the absolute value of) the local excess  entropy   in the liquid  layers closest to the wall  which induces the observed  diffusion enhancement in these layers. These results thereby show that  the application scope  of the scaling law which has so far only been used for the description of the bulk liquid diffusion can be extended to the   diffusion in liquids under nano.scale confinement.

%The natural hydrophobicity of rare-earth oxides is intriguing and yet not fully understood. In this ab-initio molecular dynamics investigation we reveal a new origin of hydrophobicity on \ce{CeO2} (100) surface. We demonstrate that such surface induces a densely packed water layer above it that behaves like an hydrophobic surface for the remaining liquid water above. We further show that this effect is measurable by the lateral liquid water diffusion, which results enhanced with respect the standard value of bulk water for the same thermodynamic conditions.

%This opens the door for studying water hydrophobicity taking into account a new paradigm: hydrophobicity triggered by specific water patterns.

\end{abstract}

\maketitle

Diffusion is a fundamental and the most ubiquitous property of the liquid state, but its many aspects  still elude a comprehensive understanding  on the microscopic level. The statistical theory is still unable to predict the rate of diffusion in a liquid based on the information on the interparticle interaction forces. One of the intriguing phenomena posing a problem for the liquid theory is the  enhancement of the  diffusion in the proximity of a non-interactive confining surface that was observed in many liquids of widely different nature, from water to liquid noble gases\cite{stanley, scheidler, agosta1, drossel, cicero}. 

Structural and dynamical properties of liquids confined in nanometer-sized channels or pores are known to be different from those of the respective bulk liquids. In recent years considerable research efforts has been focused on investigating the changes of the properties of water  upon nanoscale confinement. Understanding  these changes is of significant interest for  a wide range of applications, e.g., biological systems, flow in organic and inorganic media, the properties of materials such as zeolites and cements, and nanoparticles in solution. An important question concerns the influence of the properties of the confining materials. A fundamental question is when the dynamical properties of confined  water can be regarded as independent of confining surfaces or the fact of confinement always has  the dominant influence on the nanoscale liquid. It is established, both in real experiments and the atomistic simulations that the hydrophilic confining surfaces significantly slow down the dynamics of confined  water within several  adjacent layers. On the other hand, it was found that the proximity of a hydrophobic confining surface results in a measurable enhancement of the diffusion and other dynamical properties of water  \cite{agosta1, agosta2}. 

A similar effect of the diffusion  enhancement under nanoscopic-scale confinement  was also observed in a wide range of other liquids,  from polymers to simple atomic systems, which indicates the universal nature of this phenomenon. So far, no comprehensive unifying explanation of these puzzling observations has been proposed. In the case of water, it was suggested \cite{stanley} that the enhanced molecular mobility can be a result of the reduction of the number of the hydrogen bonds due to the presence of a neutral hydrophobic surface. This conjecture, however, cannot explain the fact that  a similar enhancement  effect  was observed  in liquids composed of particles with spherically symmetric interactions near the smooth neutral or repulsive surface \cite{scheidler}. It is obvious that a more general mechanism is needed to explain the universality of the observed enhancement, and the most general mechanism that can be conceived must be based on purely geometric considerations.

In this paper, we suggest such a mechanism. We report a molecular dynamics simulation of a simple liquid under confinement by a smooth neutral wall. We observe a significant enhancement of the lateral diffusion in the layers adjacent to the wall.   Evidence is presented that that the observed diffusion enhancement near the neutral smooth repulsive wall is of purely entropic origin. We demonstrate that the enhancement effect can be quantitatively accounted for using an earlier reported scaling law for liquid diffusion \cite{dzugutov1,dzugutov2} that relates the diffusion rate to the local excess entropy. The results of this study also show that this law which has earlier only tested for to describe the diffusion in bulk liquids can be successfully applied to the description the diffusion in liquids under nanoscale confinement.

We start with recalling the fundamental connection between the dynamical properties of dense liquids and the liquid structure.  The structural relaxation and transport properties of dense liquids are dominated by the mechanism called ``cage diffusion'', This term was coined by E.G.D Cohen \cite{cohen} who recognised it in connection with the specific feature of the spectra neutron scattered on simple dense liquids, the so.called de Gennes narrowing \cite{degennes} which refers to the reduction of the half-width of the spectra around the wave vector $Q–0$ corresponding to the position of the main peak of the structure factor $S(Q)$ \cite{hansen}. This wavelength manifests the dominant periodicity of the local order in dense liquids with the period  which corresponds to the average nearest-neighbour distance. This structure can be viewed as a local order represented by the cage of nearest neighbours in which every particle remains confined for a measurable time. The elementary diffusive motion of a particle can only be possible as a result of the cage rearrangement. In this way, the diffusion dynamics  in dense liquids is fundamentally coupled with the local structural relaxation dynamics. On the other hand, the effect of  de Gennes narrowing demonstrates that  the relaxation dynamics is controlled by the liquid structure manifested by the main peak of structure factor $S(Q)$, the latter being the Fourier-transform of the density radial distribution  function $g(r)$ \cite{hansen}. 

The proximity of a confining surface significantly reduce the  number of neighbours involved in the local structural environment  of a liquid particle. As concluded above it, this is also expected to  affect the liquid structural relaxation and, thereby, the diffusion rate.   Now, we have the following problem. The liquid structure is expressed  by either the structure factor or the radial distribution function. On the other hand, the diffusion coefficient is a number.  In order to construct a theoretical model connecting the self-diffusion coefficient  in a liquid to its structure we have to quantify the latter with a single quantity. In general, the structure of a liquid is a full set of the correlation functions describing the mutual arrangement of its particles. We therefore need to use  a functional of these correlation functions that would be possible to relate do the diffusion coefficient. This functional  is the excess entropy  $S_{ex}=S-S_{pg}$, where $S$ is the system's total entropy and $S_{pg}$ is the entropy of the perfect gas \cite{evans, raveche, mountain, wallace}. 

The process of structural relaxation in a liquid can be viewed as a random walk in the 3N-dimensional space of its configurations where $N$ is the number of particles. It is assumed that an elementary step in this process is a local particle rearrangement moving the system to a new configuration point. It only happens if the destination configuration is open the probability of which can be expressed through the excess entropy as $e^{S_{ex}}$. Therefore, the diffusion coefficient is supposed to scale universally as  \cite{dzugutov1,dzugutov2}

\begin{equation} 
 D = D_0 e^{s_{ex}}
\end{equation}

The excess entropy can be expanded in terms of the contributions from  n-body correlation functions as

\begin{equation} 
s_{ex} = s - s_{pg} = \sum_{n=2}s_n
\end{equation}

We note that the kinetic theories of liquid state are essentially two-body theories based on the hard-sphere model approximation \cite{cohen}. Therefore we assume that the excess entropy in Eq 2 can be confined to the two-body approximation. For en isotropic liquid environment two-body correlation function can be reduced to  the spherically symmetric  radial distribution function $g(r)$ \cite{hansen}, and the respective two-body excess entropy approximation  can be expressed as  \cite{evans, raveche, mountain, wallace}

\begin{equation} 
 s_2= -2 \pi \rho \int_0 ^\infty \{ g(r) \ln [g(r)] - [g(r)-1] \} r^2 dr
\end{equation}

With this form of the excess entropy, and using the time scaling in terms of collision frequency the scaling law Eq.2 can universally relate the liquid diffusion coefficient to the liquid structure \cite{dzugutov1,dzugutov2}. This universal scaling law  was found to be successful in describing  diffusion in a wide variety of liquids \cite{yokoyama, babak,pnas}. Here we shall use it to analyse the phenomenon of the enhancement of liquid dynamics near a confining surface.

First, we have to estimate how the local excess entropy $s_3$ is changed by the proximity of the non-interactive confining surface. For that purpose we have to modify the Eq.2 in order to take into account the loss of correlations truncated by the wall. The area of the segment of a sphere   of radius $r$   truncated by a plane  at the distance $d$ from its centre is

\begin{equation} 
 A = 2 \pi r (r+d)
\end{equation}

%and its share of the total  area of the sphere is

%\begin{equation} 
% f(r,d) = A/4\pi r^2 =    (r-d) /2 r
%\end{equation}
%\begin{figure}[h!] %figure 1
%\makebox[\textwidth][c]{\includegraphics[width=1.1\textwidth]{plot_msd.pdf}}%
%\includegraphics[width=1.1\columnwidth] {plot_msd.pdf}
%5\caption{Mean-square displacements in the layers and in the bulk.}

Accordingly, the part of $s_2$ which represents the correlations of the central particle with the particles that happened to be within the segment of the  sphere cut by the wall can be estimated, for $r>d$, as:

\begin{equation} 
s_2(d) = - \pi \rho \int_0 ^\infty \{ g(r) \ln [g(r)] - [g(r)-1] \} r(r+d)dr
\end{equation}

the entropy gain can then be evaluated:

\begin{equation} 
\Delta s_2(d) = s_2(d)-s_2
\end{equation}

According to the scaling law (2) this reduction of the absolute value  of $s_2$  due to the loss of correlation beyond the confining wall is expected to result in the   enhancement of the diffusion rate for the particles near the wall which can be estimated as
 
\begin{equation} 
  D_d/D_b = e^{\Delta s_2}
\end{equation}

where $D_d$ and $D_{b}$ are the diffusion rates at the distance $d$ from the wall and in the buk liquid, respectively.

In order to test this conjecture, we performed a molecular-dynamics simulation of a dense liquid using the system of 17820 particles interacting via the Lennard Jones (LJ)  pair potential.  The simulation was carried out at the triple point  density of the LJ system $\rho=0.84$ and the temperature $T=1.0 $. We note that all the quantities we report are expressed in the LJ reduced units. The system was confined to a cubic box where the boundary conditions were assumed to be periodic in two dimensions (X and Y), and confined between two walls perpendicular to Z axis. The interaction between the constituent particles and the wall at the distance $r$ were described by the potential energy function $U(r) = k(r-r_0)^b$, where $r_0$ is the position of the wall, $k$ is the force constant and $b$ relates to the stiffness of the wall. \\

In order to investigate the effect of the confining wall on the liquid diffusion we consider layers of the simulated liquid parallel to the confining wall, with the  layer width chosen  100 \AA . In each layer we consider lateral diffusion in the plane perpendicular to Z axis. The transnational diffusion constant when an anisotropic external potential is acting to the system, like a rigid wall, is determined from the Smoluchowski equation. The components of the diffusion coefficient parallel to the external field are evaluated from the mean-square displacement corrected by the probability for a water molecule to diffuse out of a certain defined layer during the time interval. 

\begin{equation} \label{eq:4} 
  D_{xy} (s) = \frac{1}{4} \lim_{t \to \infty}
  \frac{\text{MSD}_{xy} (s, t)}{t P(s, t)} \, ,
\end{equation}

where the mean square displacement is a function of water layer thickness defined as the distance $s$ from the surface. $P(s, t)$ is the probability for a water molecule to stay in the in region defined by $s$ for a certain time $t$\cite{agosta2}.\\

%\textbf{Lorenzo,  include here the MSD plot and the description of the diffusion calculation in the layers}

\begin{figure}[h!] %figure 2
\includegraphics[width=1.0\columnwidth] {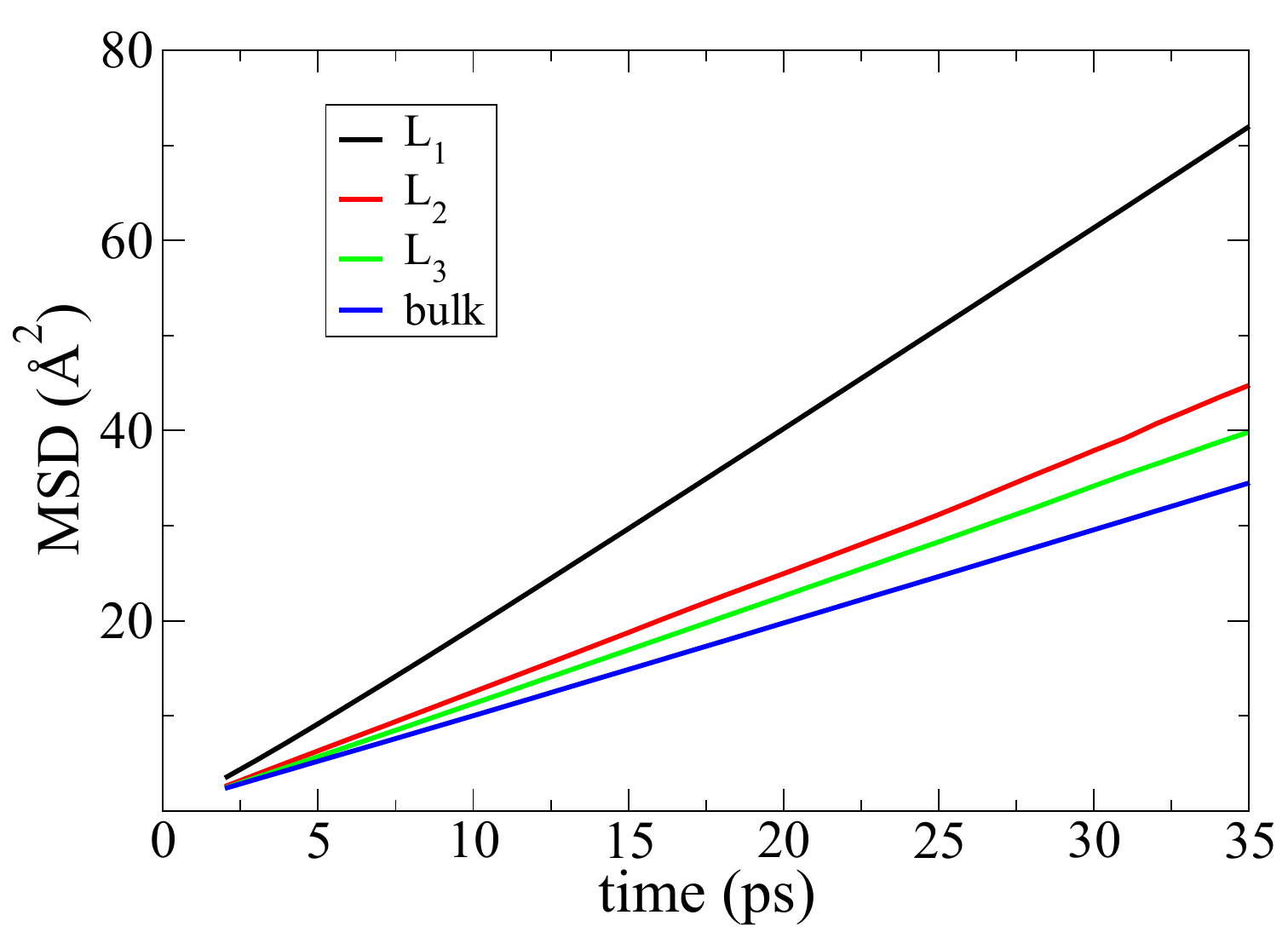}
\caption{The lateral mean square displacement. Each layer has 4 \AA \, thickness.}
\end{figure} 

\begin{figure}[h!] %figure 2
\includegraphics[width=1.0\columnwidth] {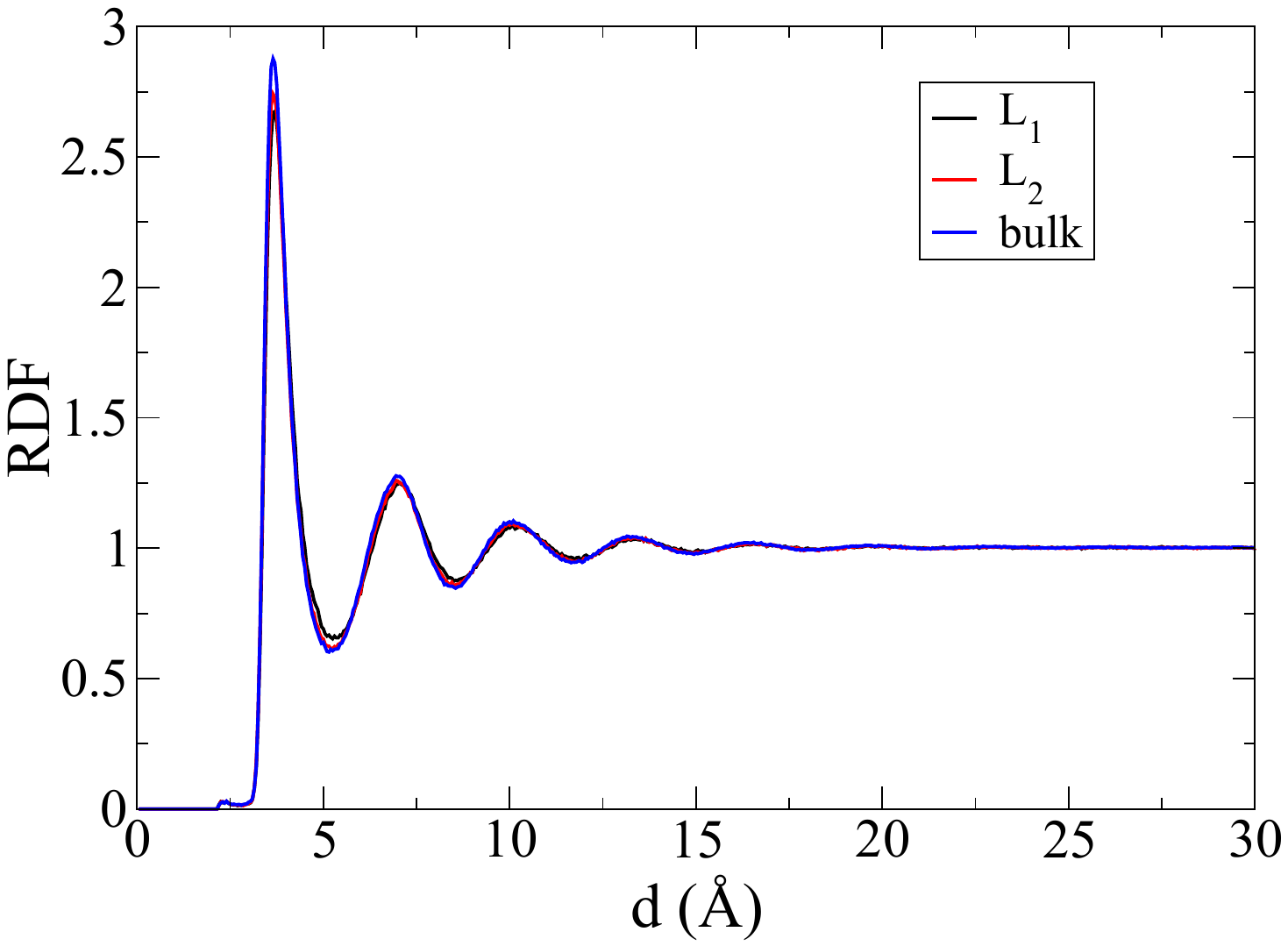}
\caption{The radial distribution functions.}
\end{figure}

\begin{figure}[h!] %figure 3
\includegraphics[width=1.0\columnwidth] {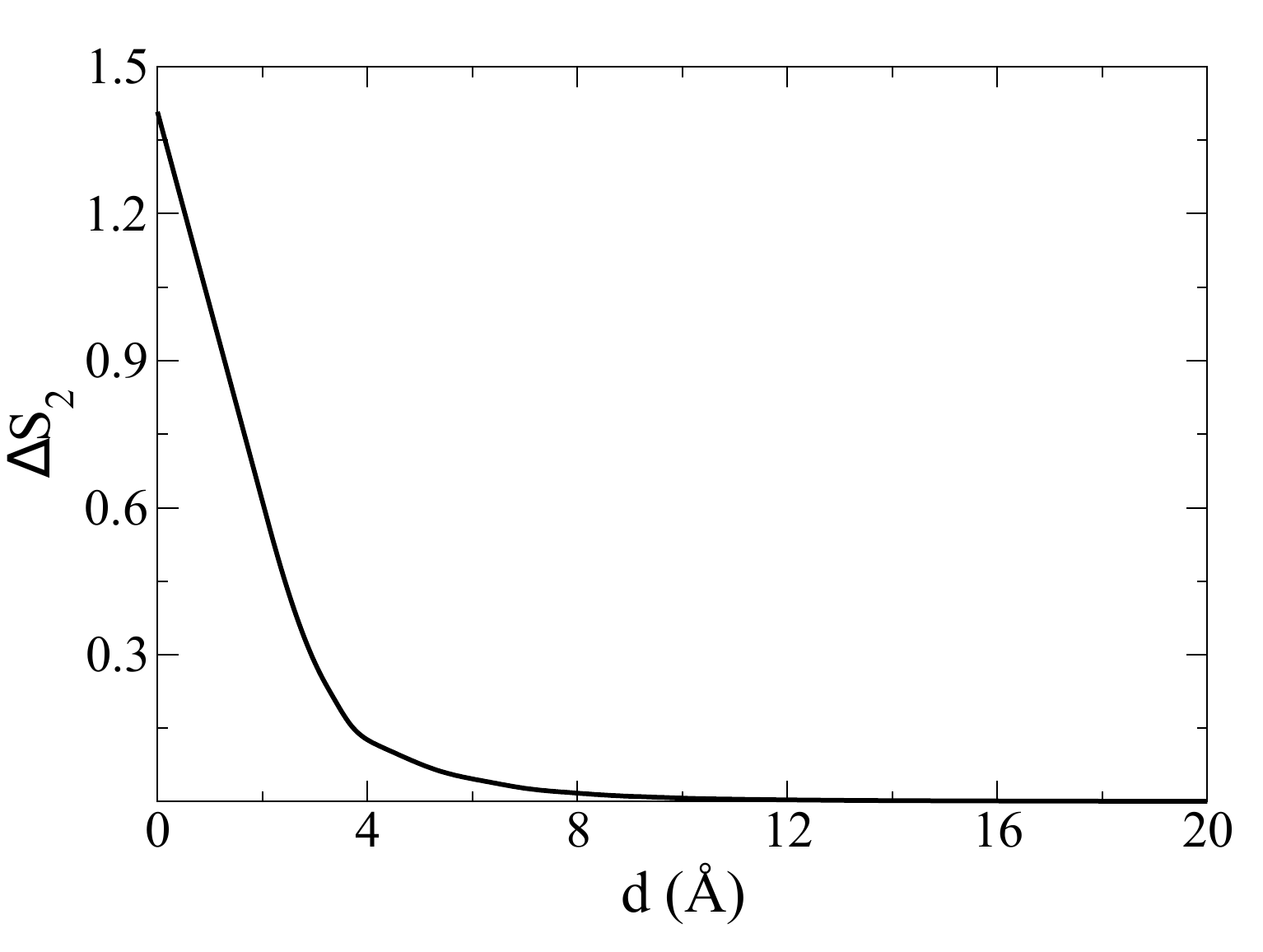}
\caption{The variation of the pair approximation of excess entropy calculated at distance $d$ from the confining wall calculated according to Eq. 6.}
\end{figure}

\begin{figure}[h!] %figure 4
\includegraphics[width=1.0\columnwidth] {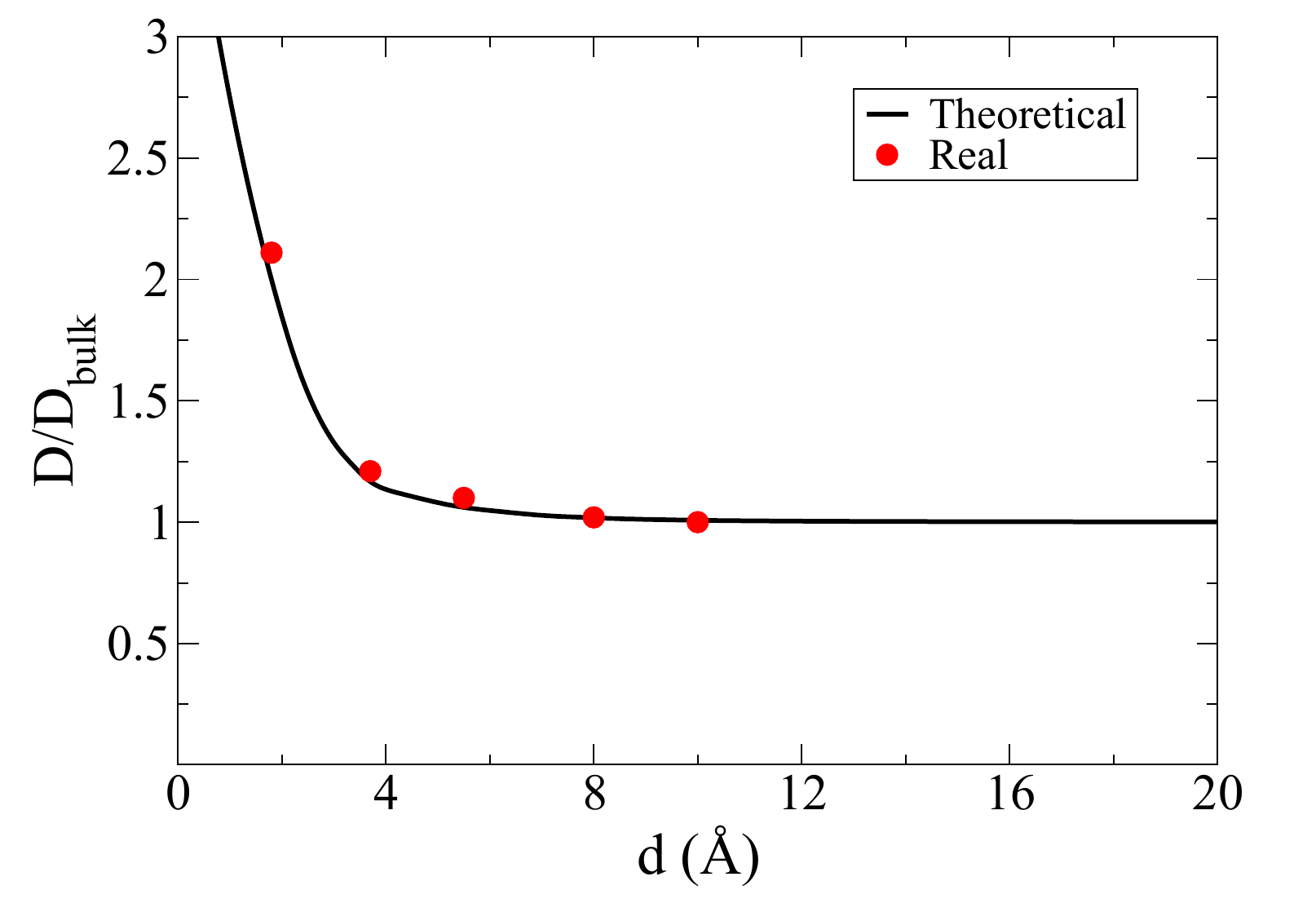}
\caption{Dots, diffusion coefficient values in the layers of simulated liquid calculated at the indicated distance from the confining wall. The curve indicates the  variation of the  diffusion coefficient as a function of the distance from the wall calculated   according to Eq. 7}
\end{figure}

In Fig.1 we compare the mean-square displacement in the first two layers next to the wall with the bulk 2D diffusion. The effect of diffusion enhancement induced by the proximity of the wall is quite apparent.

We can now use the  theoretical model presented above to estimate how presence of a confining wall can increase the diffusion coefficient $D_d$ in the liquid layer  at the distance $d$ from the wall as compared with that in the bulk liquid  $D_b$, Since, as we has shown in Fig. 2, the difference between the liquid structure close to the was is not significantly different from that in the bulk liquid, we can use the bulk $g(r)$ in the Eq. 3 to calculate the reduction of the absolute value of  excess entropy $\Delta S_2$ caused by the truncation of the correlation sphere by the confining wall. In Fig. 3 we show how this quantity depends on the distance from the wall. As one could presume from the purely geometric considerations,  the observed  effect of the proximity to wall upon $S_2$ is quite significant at the distance of the first layer, but it is quite short-ranged being limited to the first couple of layers.   

The change of the excess entropy as a function of the distance from the wall having been calculated, we can now estimate the expected enhancement of the diffusion rate according to Eq. 6, and compare it with the actual variation of the  2D lateral diffusion rate, calculated using eq. 8,  within layers at different distances from the wall. 

%The diffusion coefficients have been calculated from the asymptotically linear behaviour of the respective mean-square displacement  $<r_2(t)$ using the Einstein relation for two-dimensional diffusion:

 These results are presented in Fig. 4. We can observe quite good agreement between the actual diffusion rate and the theoretical prediction based on the excess entropy scaling.

We can also estimate the error in the theoretical prediction of the diffusion enhancement according to the entropy scaling relation that arises from the reduction of the full excess entropy $s_{ex}$ to its two-body approximation $s_2$. For that purpose we can use the results reported in Ref. \cite{evans}.  These authors estimated the share of the  contributions to the excess entropy  beyond the the two-body approximation in the LJ liquid around its triple point. It was found to be about $7\%$.

%Respectively, the same estimation of the higher-order contribution is applicable to $\Delta s$. Then the error we introduce in the calculation of  $D_d/D_b$, Eq. 7, for the closest layer to the wall is about 5%.  

Some concluding remarks are in order.\\

1. The theoretical model interpreting the diffusion enhancement in liquids near confining walls in terms of the excess entropy scaling law can be easily extended to explain similar diffusion anomalies in liquids near other kinds of surfaces, including the surface separation the coexisting  liquid and gas  phases. 

2. It is quite clear from the geometrical arguments which were presented above that the diffusion enhancement effect in liquids induced by the proximity of a confining surface is expected to be stronger if the surface shape is concave with the curvature radius of nanometer scale. This conjecture is of significance for the applications like water dynamics in nanoporous materials.

3. The results presented here can be regarded as another successful test of the scaling law relating  liquid diffusion to the excess entropy. The results demonstrate that application scope of this law can be extended to include the description of the liquid diffusion in confined geometries.

In summary, we performed a molecular dynamics simulation of a confined liquid  demonstrating  a significant diffusion enhancement in the layers near the confining walls. It is shown that this effect can be successfully accounted for on quantitative level using the entropy scaling law for the liquid diffusion. The observed  diffusion enhancement is demonstrated to be a result of the reduction of (the absolute value of) the local excess entropy caused by the truncation of the structural correlation sphere of a diffusing particle by the confinement. A conceptually significant conclusion of these results is that the scaling law we successfully tested here can be extended to the description of the diffusion in confined liquids. 

The authors gratefully acknowledge the numerous very useful and illuminating discussions with Prof.  Dr. Babak Sadigh.\\

\end{document}